\documentclass[prl,superscriptaddress,showpacs,twocolumn]{revtex4}
\usepackage{amsmath,amsfonts,amssymb}
\usepackage{graphicx}

\begin{document}

\title{Dynamical and Statistical Criticality  in a Model of Neural Tissue}

\author{Marcelo O. Magnasco}
\affiliation{Laboratory of Mathematical Physics, The Rockefeller
University, 10021 New York, NY USA}

\author{Oreste Piro}
\affiliation{Departament de F'sica and IFISC(CSIC-UIB), Universitat de les Illes Balears, 07122 Palma de Mallorca, Spain.}

\author{Guillermo A. Cecchi}
\affiliation{Computational Biology Center, T.J. Watson IBM Research
Laboratory, 1101 Kitchawan Rd., Yorktown Heights, NY USA}

\begin{abstract}
For the nervous system to work at all, a delicate balance of excitation and inhibition must be
achieved. However, when such a balance is sought by global
strategies, only few modes remain balanced close to instability, and
all other modes are strongly stable. Here we present a simple model
of neural tissue in which this balance is sought locally by neurons
following `anti-Hebbian' behavior: {\sl all} degrees of freedom
achieve a close balance of excitation and inhibition and become
``critical'' in the dynamical sense. At long timescales, the modes
of our model oscillate around the instability line, so an extremely
complex ``breakout'' dynamics ensues in which different modes of the
system oscillate between prominence and extinction. We show the system develops various anomalous statistical
behaviours and hence becomes self-organized critical in the statistical sense.  \end{abstract}


\maketitle

Dynamical systems theory holds that systems of interest should be
{\em structurally stable}: their behavior should not drastically
change with small perturbations of the defining dynamics
\cite{Gucken}. Thus high-order criticality, the {\em simultaneous}
presence of several critical features such as Hopf bifurcations, is
not expected to be ever observed in a natural system. However
natural systems lacking such structural stability are not
infrequent: within neuroscience examples include {\em dynamically
critical} systems such as line attractors \cite{lineattractors} in
motor control \cite{motorcontrol} and decision making
\cite{decisionmaking}, and self-tuned Hopf bifurcations in the
auditory periphery \cite{hopfear} and olfactory system
\cite{freeman}. There are also examples in neuroscience of {\sl
statistical criticality}~\cite{Per}: spontaneous heavy-tailed or
scale-free fluctuations typical of critical phase transitions, such
as neuronal avalanches in cortical slices \cite{avalanches},
anomalous correlations in the retina \cite{schneidman} and in
functional imaging \cite{corrfmri}, and models based on
simulations of the highly non-linear dynamics of spiking elements,
display avalanche-like statistical criticality \cite{Chen,Theo}.
There is no real understanding
of a relation between these different concepts of
criticality; developed turbulence, a well-studied example, displays
both statistical criticality \cite{TurbScaling} 
and dynamical criticality (extensive number of zero Lyapunovs \cite{Goy}), but a relationship
between them is far from clear.

We present a simple model of {\em neural tissue}, an
anti-Hebbian network which constantly forgets; this network
spontaneously poises itself at a dynamically critical state in which
an extensive number of degrees of freedom approach Hopf
bifurcations, becoming arbitrarily sensitive to external
perturbations. As the dynamics controlling this state has itself a
marginal fixed point, the eigenvalues do not converge but fluctuate,
close to the imaginary axis; when they become slightly unstable, the
corresponding mode ``breaks out'' and becomes more prominent, and as
they become slightly stable the mode slowly damps out. This breakout
dynamics displays avalanche-like activity bursts whose sizes are
power-law distributed. Within these epochs the neurons of our model
are slightly correlated; yet, as the number of small but significant
correlations is high, the model has strongly correlated network
states \cite{schneidman}. This system is, on the short time-scale,
sensitive in bulk to any outside input, even if applied only to a
small subset of the neurons; however, it does not learn. In fact,
being anti-Hebbian, it constantly forgets. We show that we can
enrich the dynamics adding, to the term which is anti-Hebbian respect to regular
correlations,  another term ``positively'' Hebbian to directed correlations,
i.e., those causal in the sense of Granger \cite{Granger}.  
Then the network may learn ``predictable'' stimuli, yet will
stay unable to learn noise, and  will display timing-dependent
synaptic changes reminiscent of spike-timing dependent plasticity (STDP, \cite{STDP}).

\begin{figure}[tb] 
\includegraphics[width=3.2in]{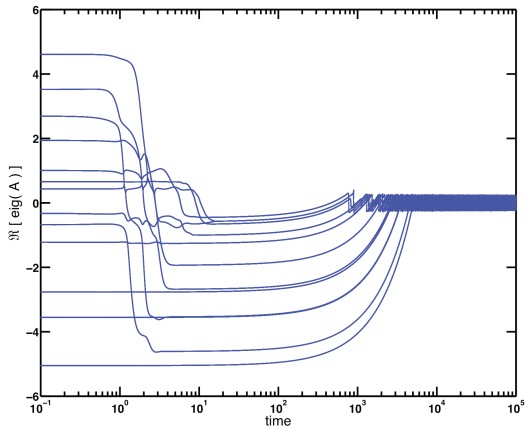}
\caption{Relaxation of the real parts of the eigenvalues of $A$. For
clarity of illustration, $N=20$. At short times ($\approx 1$) all
eigenvalues with positive real parts relax to having negative real
parts; they typically overshoot and flip sign in doing so. On a
scale given by $\alpha=10^{-3}$, all real parts relax to the
vicinity of the real axis. Beyond this scale, all eigenvalues
fluctuate around the real axis. } \label{fig1}
\end{figure}

We now present our model. The activities of a set of neurons,
encoded in the vector $\bf x$, evolve 
under the synaptic connectivity matrix $A$; meanwhile $A$ itself
evolves, at a slower pace $\alpha$, under an anti-Hebbian rule.
\begin{equation}\label{EQ1} \dot {\bf x} = A {\bf x}
\end{equation}
\begin{equation}\label{EQ2}
\dot A = \alpha ( I - {\bf x x^\top })
\end{equation}
where the matrix $A$ encodes the synaptic connections, $\alpha$ is
the speed of synaptic evolution, assumed slow, and $I$ the identity
matrix. Inputs ${\bf i}(t)$, neuronal noise $\xi(t)$, and nonlinear limiting terms such as ${\bf x}^3$ would normally be added to the RHS of eq (1) but that shall not be necessary for now. 
From eq. (2) the matrix $A$ would stop evolving when the components of $\bf x$
have unit variance and are uncorrelated to one another. 

The evolution of this system is
surprisingly complex and generates several different timescales, as
shown in Figures 1 and 2. For a random initial $A$, first, the eigenmode
$\bf e$ having the eigenvalue with the largest positive real part
starts to diverge, and as it does so, incurs a large penalty $\dot A
\approx -\alpha {\bf e e^\top}$. This happens on a timescale of
order 1 (Figure 1). After all eigenvalues are to the left of the imaginary
axis, a second dynamical regime ensues in which the real part of
eigenvalues increases at a rate $\alpha$ until the real part
approaches zero. Finally, the eigenvalues migrating to a strip
around the imaginary axis, but instead of relaxing to this value,
they oscillate around their equilibrium positions (Figure 2)

\begin{figure}[t]
\includegraphics[width=3.2in]{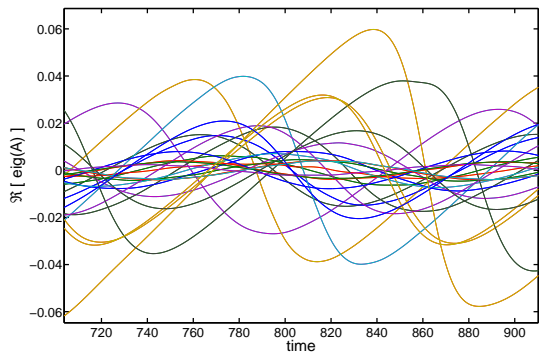}
\caption{Zooming in the rightmost portion of Fig. \ref{fig1}. 
Starting from a fixed point of Eq.~(\ref{EQ4}), i.e., an antisymmetric matrix, the
eigenvalues fluctuate around the instability line with a timescale $\approx \sqrt{ \alpha}$ } \label{fig2}
\end{figure}

It is illustrative to use a long-time
approximation ($\alpha<<1$) when we add a noise source to each neuron:
\begin{equation}\label{EQ3} \dot {\bf x} = A {\bf x} + {\bf \xi} (t)
\end{equation}
where $\bf \xi$ is white noise, assumed small: $\langle \xi_i(t)
\xi_j(s)\rangle = 2kT \delta_{ij} \delta(t-s)$ with $kT \ll 1$.
First we write Eq.~(\ref{EQ3}) in the basis of the eigenvectors of
$A$, where each component becomes an Ornstein-Uhlenbeck process (OUP) with
(complex) decay rate $\lambda_i$, and the amplitude of the OUP becomes $-kT/\Re (\lambda)$ when $\Re (\lambda)<0$,
and divergent otherwise. Then we compute $\langle {\bf x x^\top }|A
\rangle$, the expectation value of the correlation of the $\bf x$
under the assumption that $A$ is constant, and use this value in
Eq.~(\ref{EQ2}):
\begin{equation}\label{EQ4}
\dot A = \alpha ( I - \langle {\bf x x^\top }|A\rangle )
\end{equation}
Diagonalize  $AV = V\Lambda $ where $V$ are the right
eigenvectors and $\Lambda $ the diagonal matrix of eigenvalues
$\lambda_i$. Define 
$$\dot A = \alpha(I - V B \overline V) $$ 
The elements of $B$ are given by $$ B_{ij}={(
V^{-1}\overline{V^-1})_{ij} \over \lambda_i+\lambda^*_j }$$ 
whose diagonal elements are  $1/2\Re \lambda $; $V B \overline V$ is
related to the inverse of the matrix $V\Re\Lambda V^{-1}$ (which we
may call the ``real part'' of $A$). If $A$ had orthogonal
eigenvectors, then $V^{-1}=\overline V$ and hence $B$ would be the
diagonal matrix having $1/(2 \Re\lambda)$ in the diagonal, from
where in the steady state $\dot A=0$ we would obtain
$kT  = \Re\lambda $. 

Because $\dot A$ is
symmetric, $A$ evolves as $A(0)+S(t)$ where $S$ is a
symmetric matrix; the antisymmetric part of $A$ is a constant of the motion.  The evolution annihilates the symmetric part of $A$ until the only piece left is the identity matrix multiplied by $-kT$, which is needed so the amplitudes of the OUPs is1. Therefore
the fixed points of Eq.~(\ref{EQ4}) has the form
$A^*={1\over 2} (A(0)-A^\top(0)) - kT \mathbb{I}$, i.e., the
antisymmetric component of the starting matrix, minus $kT$ times the
identity. 

\begin{figure*}[tbhp] 
\includegraphics[width=7in]{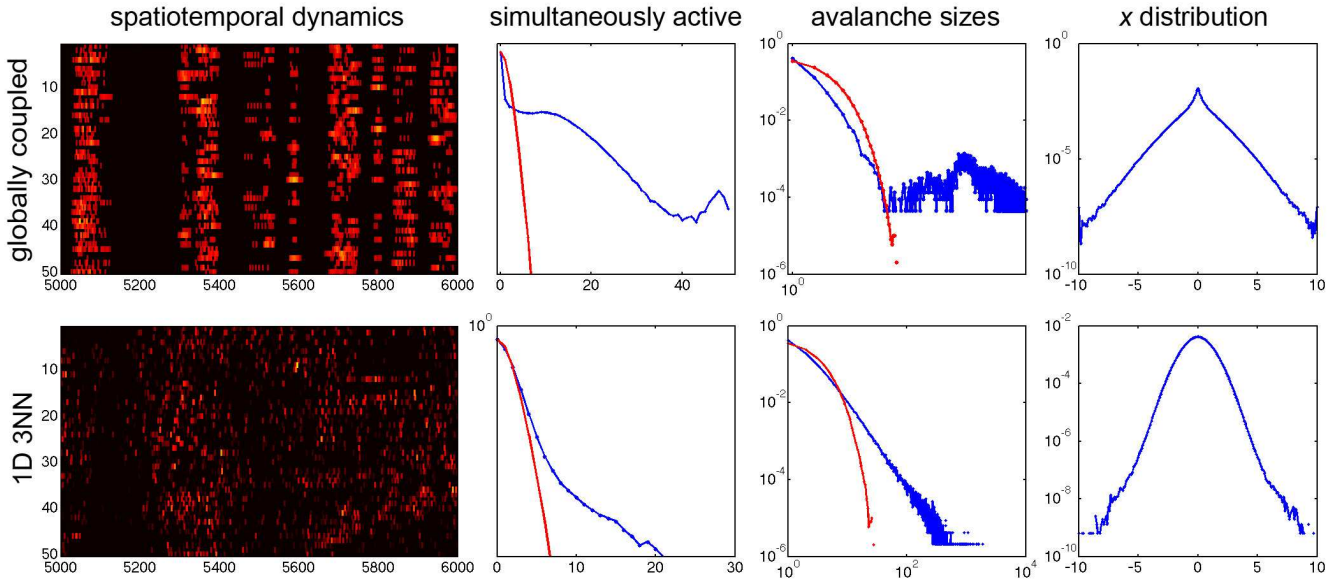}
\caption{Statistical criticality in our model.  Top row, globally coupled (unrestricted $A$). Bottom row, nodes arranged in one dimension with periodic boundary conditions; only entries of $A$ up to third nearest neighbour are allowed to be nonzero.
First column, a display of the spationtemporal dynamics. Second column: the distribution of the number of simultaneously active units in the dynamics (blue) and in surrogate data (red); compare to \cite{schneidman}. Third column, sizes of avalanches (blue), vs. surrogate data (red); note in the 1D case the power-law distribution of avalanche sizes, while the globally coupled ($\infty$-D) case shows a piece of a power-law followed by a large lump of rather large avalanches (as clearly visible in the spatiotemporal plot). Fourth column, marginal distribution of the values of $x$ (invariant under surrogation).  } \label{fig3}
\end{figure*}

However, this fixed point turns out to be only marginally stable.
This is easily verified: take a time derivative of Eq.~\ref{EQ2}
\begin{equation}\label{EQ5}
\ddot A = {d\over dt} \dot A = -\alpha {d \over dt} {\bf x x^\top }
= -\alpha ({\bf \dot x x^\top + x {\dot x}^\top }) \end{equation}
Using $\dot {\bf x} = A {\bf x} + {\bf \xi}(t) $ and then
reasserting that at the fixed point $I = {\bf x x^\top } $
\begin{equation}\label{EQ6}
\ddot A = -\alpha (A+A^\top)
\end{equation}
from where we see the precise kind of marginal stability in
question: the symetric component of the matrix $A$ follows an undamped harmonic oscillator equation. 
So {\em this fixed point is itself a multidimensional Hopf bifurcation}. 
The real parts of the eigenvalues of $A$ oscillate
around the instability boundary with a frequency $\sqrt{2\alpha}$.
There is therefore a {\em new timescale} given by $2\pi\over
\sqrt{2\alpha}$ (Fig.~\ref{fig2}). White noise such as in
Eq.~(\ref{EQ3}) may decohere the dynamics a bit, but is not necessary
for driving it; even if the noise is turned off entirely, the system
continues to oscillate. The oscillation frequency $\sqrt{2\alpha}$
is approximately the
geometric mean between the neuronal oscillation timescale (in this
Letter, $\approx 1$) and the synaptic update timescale $\alpha$. In
a real neuron, the oscillatory timescale is bound to be in the
10-120 Hz frequency bands, while the synaptic update timescale would
be in the several minutes. The geometric mean between these, marking
the scale in which any given mode would spontaneously
activate and deactivate, would be in the seconds range,
bridging these two scales; this timescale marks the ability of our model to ``switch task''
and corresponds roughly to the timescale of thought. It may seem counter-intuitive
that such a fast timescale would be a consequence of slower synaptic
update, unless one realizes that only in conjunction with the {\em
population} dynamics close to a {\em dynamically critical} state,
minute alterations in synaptic strength across a population of
neurons may change dynamical behavior in a mere fraction of the time
required to swing a single synapse from low to high strength.

The largest oscillations in Fig.~\ref{fig2} show an asymmetry that
is interesting to explain. Consider our equations in the simple
noiseless case of a single degree of freedom:
\begin{eqnarray}\label{EQ7}
\dot x & = & \lambda x \label{EQ7} \\
\dot \lambda & = & \alpha ( 1 - x^2 ) \label{EQ8}
\end{eqnarray}
The system has two marginally stable fixed points ($x=\pm 1$),
responsible for the oscillations. Excursions on the first quadrant,
left of $x=1$, produce runaway behavior of $x$ and increase in
$\lambda$, until the quadratic term in Eq.~(\ref{EQ8}) kicks in
before $\lambda$ turns negative, leading to different slopes in the
up- and down-swings. This one-dimensional example also illustrates
the origin of the statistical criticality shown in Fig.~\ref{fig3}
{\sl in the absence} of driving noise: the marginal stability of the
modes and the mismatch of the dynamical and learning time-scales
interact to create a long-tailed distribution of activity and
avalanche sizes.

In an attractor neural net, such as a Hopfield net, the antisymmetric components of $A$ are either null or small, and learning is carried out by using a Hebbian rule, which then encodes the learned objects in the symmetric part of $A$. In our case, our anti-Hebbian dynamics takes control of the symmetric part of $A$ and uses it to create the critical, highly plastic state we have described. The antisymmetric component of $A$ is evidently untouched by Eq. (2), an invariant of the motion, and is the only degree of freedom available for learning in our system. 
The $\bf x x^\top$ term is an instantaneous density of correlation, which eq (2) integrates in time due to the smallness of $\alpha$; let us write it as 
\begin{equation} ({\bf xx^\top})_{ij} = \int \delta(s) x_i(t+s) x_j(t-s) ds \label{symcorr}
\end{equation}
An antisymmetric analog of this correlation density in the RHS of Eq. (2) would be given by partial directed correlations, correlation functions which attempt to isolate influences between time-series embodying Granger causality \cite{Granger}. Such correlation functions are obtained through a kernel which is antisymmetric in time, causing the correlation density to become antisymmetric in the neuron indexes; the simplest analog of Eq. \ref{symcorr} would be to replace the $\delta(s)$ by the Hilbert transform $1/s$ to obtain
\begin{equation}
C_{ij} = \int { x_i(t+s) x_j(t-s) \over s } ds 
\end{equation}
The Hilbert transform kernel $1/s$ is divergent at both short and long timescales, and should be both ultraviolet and infrared cutoff according to the fastest and longest timescales which the system can use for its evaluation; the fast timescale controls the transition between increasing synaptic strength when the presynaptic neuron leads the postsynaptic one, to decreasing it in the opposite case, and reflects the accuracy with which the system can compute simultaneity. The slow timescale controls how much memory is kept of previous activity, i.e., over which range time intervals the pre- and post- synaptic activities are evaluated. When the Hilbert kernel is cutoff according to these two timescales, the synaptic rule left looks precisely like STDP \cite{STDP}

Our results suggest a different light in which multi-electrode
recordings may  be fruitfully looked at. Our model proposes a view
of neuronal tissue as showing coexistence and superposition of
different modes of neuronal activity, which can be simultaneously
long-lived in terms of the timescales of electrical activity, yet
extremely fast in terms of synaptic update timescales. The
fundamental distinguishing factor between each of the activated
modes is the different phase relationship of each neuron with
respect to the underlying oscillation. Analysis methods which aim to
tease apart these epochs of behavior can be devised by understanding that the different 
modes are distinguished from one another by looking at the activity of single units 
in a coordinate frame constructed from the activity of the other units, rather than of 
external references. Finally it is worthwhile to remark that, since the dynamics consists of the activation and deactivation of modes of behaviour given by eigenvectors which are in general delocalized, the dynamics of our net is resilient to stochasticity or even failure in individual units. Detailed analysis of this resilience shall be carried out elsewhere. Similarly, because the dynamical modes are delocalized, the system is sensitive to the topological structure of the underlying network on scales much longer than individual connections or plaquettes. This extended spatial sensitivity mirrors the extended temporal behaviour discussed above and will be explored elsewhere.

We have presented a simple model of ``neural tissue'', in which an underlying anti-Hebbian dynamics
permits the system to use the symmetric components of its synaptic connectivity to poise itself at 
a dynamically critical state and becomes infinitely susceptible to input which, once applied, can reverberate for long times. In the absence of inputs, this state evolves by the eigenvalues oscillating around 
the stability line, so different modes (eigenvectors) break out and then extinguish haphazardly, with a timescale which bridges the electrical and synaptic timescales.  We have
shown that learning can be encoded in the anti-symmetric component of the synaptic connectivity, driven 
by a term anti-symmetric both in space as well as time---only inputs which are Granger-causal and time-symmetry broken can be learned by this system. We have analyzed the statistics of our system to show that it can generate anomalous, heavy tailed distributions, as well as power-law avalanches, showing explicitly a connection between criticality in the dynamical and statistical senses. Finally, our model is intended to provide a scaffold to explore the implications of the reverberating circuit theory introduced by Lorente de N\'{o} and furthered by Lashley and Hebb \cite{Orbach} which, for all their influence in physiology and behavior science, have not found consistent formal expressions.
Supported in part by MCI project CGL2008-06245-C02-02 and CSIC intramural project HIELOCRIS (OP).

\vfill\eject
\appendix
\section*{Appendix (Supplementary Materials)}
Following the definitions of Eq.~(\ref{EQ3}), let us decompose $A$
as $AV = V\Lambda $ where $V$ are the right eigenvectors and
$\Lambda $ the diagonal matrix of eigenvalues $\lambda_i$. Let us
note that we can define a real matrix all of whose eigenvalues are
purely imaginary by doing $H = A - V(\Re\Lambda)V^{-1} $, this will
be useful later.

Multiplying (3) by $V^{-1}$ on the left, and defining ${\bf y}(t) =
V^{-1}{\bf x}(t)$ and $\eta(t) = V^{-1} \xi(t)$
\begin{equation} \dot {\bf y} = \Lambda {\bf y} + \eta(t)
\end{equation}
which has the solution
\begin{equation}
{\bf y}(t) = \int_{-\infty}^t e^{\Lambda(t-t')} \eta(t') dt'
\end{equation}
from where $$ \langle xx^\top\rangle=\langle x\bar x\rangle=$$
\begin{equation}\int \int_{-\infty}^t V e^{\Lambda(t-t')}V^{-1} 2kT I
\delta(t'-t'') {\overline V^{-1}} e^{\bar \Lambda(t-t'') }
{\overline V} dt' dt''
\end{equation} thus $kT/2$ times
\begin{equation}\int_{-\infty}^t
V e^{\Lambda(t-t')}V^{-1}
 {\overline V^{-1}}
e^{\bar \Lambda(t-t'') } {\overline V} dt' dt'' \end{equation} In
coordinates
\begin{equation}\int_{-\infty}
v_{ij} e^{\lambda_j(t-t')} \delta_{jk} v^{-1}_{kl}
v^{-1*}_{ml}e^{\lambda^*_m(t-t')}\delta_{mn}v^*_{pn} dt'
\end{equation} summed over $jklmn$. As the integral becomes
\begin{equation}
\int_{-\infty}^t e^{(\lambda_k+\lambda^*_m)(t-t')} = {1\over \lambda_k+\lambda^*_m}
\end{equation}
we eventually get
\begin{equation} v_{ij} \delta_{jk} v^{-1}_{kl}
v^{-1*}_{ml} \delta_{mn}v^*_{pn} \over \lambda_k+\lambda^*_m
\end{equation} further simplified to
\begin{equation} a_{im} =\sum_{jkl} { v_{ij}v^{-1}_{jk}
v^{-1*}_{lk} v^*_{ml} \over \lambda_j+\lambda^*_l }\end{equation} or
in other words,
\begin{equation}
B_{ij}={( V^{-1}\overline{V^-1})_{ij} \over \lambda_i+\lambda^*_j }
\end{equation}
and notice that it becomes $1/2\Re \lambda $ on the diagonal,
\begin{equation} \dot A = \alpha(I - V B \overline V) \end{equation}
where $V B \overline V$  is some kind of weird inverse of "Real(A)",
namely, $V\Re\Lambda V^{-1}$; also we note that if $A$ had orthogonal
eigenvectors, then $V^{-1}=\overline V$ and hence $B$ would be the
diagonal matrix having $1/(2 \Re\lambda)$ in the diagonal, from
where we'd get that for the steady state $\dot A=0$ we would get
$kT = \Re\lambda $.

\end{document}